\documentclass[preprint,12pt]{elsarticle}

\usepackage{amsmath,amssymb,bm,color}

\journal{High Energy density Physics}

\begin{document}

\begin{frontmatter}
\title{Average Atom Model based on\\
 Quantum Hyper-Netted Chain Method}
\author{Junzo Chihara\corref{cor1}}
\address{Higashi-Isikawa 1181-78, Hitachinaka, Ibaraki 312-0052, Japan}

\cortext[cor1]{E-mail address: jrchihara@nifty.com}

\date{\today}

\begin{abstract}
The study shows how to define, without any {\it ad hoc} assumption, the average ion charge $Z_{\rm I}$ in the electron-ion model for plasmas and liquid metals: this definition comes out of the condition that a plasma consisting of electrons and nuclei can be described as an electron-ion mixture. 
Based on this definition of the average ion charge, the Quantum Hyper-Netted Chain (QHNC) method 
takes account of the thermal ionization and the resonant-state contribution to the bound electrons forming an ion.

On the other hand, Blenski and Cichocki (2007) have derived a formula to determine the uniform electron density in a plasma as an electron-ion mixture by using the variational method  
with help of the local density approximation. Without use of any approximation, we derived the formula determining the electron density in an extended form on the basis of the density functional theory. This formula is shown to be valid also for the QHNC method.
\end{abstract}

\end{frontmatter}


%
\section{Introduction}
%
A liquid metal or a plasma can be taken as a mixture of electrons and ions with uniform ion density $n_0^{\rm I}$ and electron density $n_0^{\rm e}$. This binary mixture consists of ions with a definite ionic charge $Z_{\rm I}$ and the free electrons,  
interacting with each other via {\it binary} potentials $v_{ij}(r)$ 
[$i,j\!=\!{\rm I}$ or ${\rm e}$] under the charge neutrality condition $n_0^{\rm e}=Z_{\rm I}n_0^{\rm I}$. Also, the ions are assumed to behave as {\it classical} particles, while the free electrons form a quantum fluid changed into a classical fluid at high temperature. We call this mixture "the electron-ion model" for a plasma. 
Although it is important to calculate the average ion-charge $Z_{\rm I}$ in a plasma for description of thermodynamic quantities, there is no established method to determine $Z_{\rm I}$ in the electron-ion model at the present time. 
In this work, we show how to obtain the definition of the average ion-charge $Z_{\rm I}$ in the electron-ion model. Under some condition only, a plasma as an electron-nuclei mixture can be treated as an electron-ion mixture on the basis of the electron-ion model. This condition itself provides the definition of an 'average ion' in the electron-ion model. To establish this definition, we need the radial distribution functions (RDF) as already known quantities in an electron-ion mixture with the given electron-ion and ion-ion interactions, $v_{\rm eI}(r)$ and $v_{\rm II}(r)$. In this regard, the quantum hyper-netted (QHNC) method \cite{QHNC} can determine the RDFs in the mixture for arbitrary interactions $v_{ij}(r)$ [even for charged hard-core potentials].
On the other hand, Saumon and coworkers \cite{Saumon2013} have calculated $Z_{\rm I}$  with use of the QHNC method on the basis of some {\it ad hoc} assumptions about a separation of the bound electron density distribution from the total electron density distribution around a nucleus in a plasma: these assumptions can be avoided by use of the present definition of $Z_{\rm I}$. 

The QHNC method can produce the plasma structure including $Z_{\rm I}$ at arbitrary temperature $T$ and ionic density $n_0^{\rm I}$ from the atomic number $Z_{\rm A}$ as an only input \cite{PR99plas}.
It should be emphasized that the QHNC method yields structure factors in good agreement with experiments for simple metals \cite{QHLi,fpMD,CKahl98}. Also, this method can determine the electron-electron correlation in a consistent way with the ionic structure without use of jellium model for electrons \cite{JC84,JC83,JS79}.
However, the QHNC method has the following two weak points, to be improved, about the determination of $Z_{\rm I}$ and a bootstrap relation to generate the electron-electron correlation consistent with the ion structure, as follows:

\medskip
\noindent (I) In the QHNC method, the ionic charge $Z_{\rm I}\!=\!Z_{\rm A}\!-\!Z_{\rm B}$ 
is simply defined from $Z_{\rm B}\equiv \int\! n_{\rm e}^{\rm b}(r)d{\bf r}$ using the bound-electron density $n_{\rm e}^{\rm b}(r)$ for the wave equation. Therefore, this definition does not take account of the  contribution of resonant states to $Z_{\rm B}$.
On the other hand, Blenski and coworkers \cite{Blenski07a,Blenski07b,Piron2011} have derived an equation to determine the electron density $n_0^{\rm e}$ in a plasma by using the variational method (VAAQP model). Although the VAAQP model provides $Z_{\rm I}=n_0^{\rm e}/n_{0}^{\rm I}$ and $Z_{\rm B}$, it can not give  the bound-electron distribution $\rho_{\rm b}(r)$ to fulfill $Z_{\rm B}\!=\!\int \rho_{\rm b}(r)d{\bf r}$ and, thus, the ion-ion correlation $g_{\rm II}(r)$. 
In this work, we derived two relations which are valid within the framework of the density functional (DF) theory and the electron-ion model.

(a):  In the electron-ion model, the average ion charge $Z_{\rm I}$  is defined as
\begin{eqnarray}\label{e:avZi}
Z_{\rm I}&=&\frac{n_0^{\rm e}}{n_0^{\rm I}}=Z_{\rm A}-\int [n_{\rm e}(r)\!-\!n_0^{\rm e}g_{\rm eI}(r)]d{\bf r}=Z_{\rm A}-\int \rho_{\rm b}(r)d{\bf r} \label{e:DfZi0}  \\
\rho_{\rm b}(r) &\equiv& n_{\rm e}(r)\!-\!n_0^{\rm e}g_{\rm eI}(r)\,.
\end{eqnarray}
This relation is derived from the necessary condition that a plasma consisting of electrons and nuclei can be described as a mixture of electrons and ions. 
At the same time Eq.~(\ref{e:avZi}) is rewritten in the equivalent relation:
\begin{equation}\label{e:Za0}
Z_{\rm A}= \int [n_{\rm e}(r)\!-\!n_0^{\rm e}g_{\rm II}(r)]d{\bf r}\,.
\end{equation}
Here, $g_{\rm eI}(r)$ and $g_{\rm II}(r)$ are the electron-ion and ion-ion RDFs, respectively, in the electron-ion model, and $n_{\rm e}(r)$ is the electron density distribution around the nucleus, when a chosen ion in this mixture is thought as an inserted atom with a nucleus $Z_{\rm A}$.


(b): The uniform density $n_0^{\rm e}$ in the electron-ion model must satisfy the following condition:
\begin{equation}\label{e:miNn0}
\int v_{\rm es}(r)g_{\rm II}(r)d{\bf r}=\mu S_{\rm II}(0)/n_0^{\rm I}=\mu \kappa_{\rm T}/\beta \,,
\end{equation}
with $S_{\rm II}(Q)$ being the structure factor. The above equation is reduced to the result of Blenski {\it et al}, \,when  we make approximations, $g_{\rm II}(r)$ by the step-function and $S_{\rm II}(0)=0$. Equations, (\ref{e:avZi}) and (\ref{e:miNn0}), solve the problem to determine the average ion charge $Z_{\rm I}$.

\medskip
\noindent (II): The QHNC method uses the following bootstrap relation to determine the electron-electron response function $\chi_{\rm ee}(Q)$
 from the electron density distribution $n_{\rm e}(r|{\rm {e}})$ around a fixed electron in a plasma:
\begin{equation}\label{e:KOappr}
{\cal F}_Q[n_{\rm e}(r|{\rm {e}})-n_0^{\rm e}]
\equiv \int [n_{\rm e}(r|{\rm {e}})-n_0^{\rm e}]\exp(i{\bf Q}{\bf r})d{\bf r}
=\chi_{\rm ee}(Q)/\chi_Q^0-1 \,
\end{equation}
with the density response function $\chi_Q^0$ of a non-interacting system.
This relation results from the approximation used by Kukkonen and Overhauser \cite{KO,BD} for an electron gas. Since Eq.~(\ref{e:KOappr}) is an exact relation for a classical electron gas, it is appropriate in treating a high-temperature plasma. Therefore, the QHNC method with the use of Eq.~(\ref{e:KOappr}) provides a good description of the pair correlations for a hydrogen-plasma gas at low densities and high temperatures where the electrons behave as a classical electron gas \cite{ChiharaHplasma}. However, 
Eq.~(\ref{e:KOappr}) contains an approximation that the fixed electron in a liquid metal has the exchange effect to surrounding electrons: the exchange-effect part 
must be subtracted in the form:
\begin{equation}\label{e:BootStsd}
{\cal F}_Q[n_{\rm e}(r|{\rm \hat{e}})-n_0^{\rm e}]
=\chi_{\rm ee}(Q)/\chi_Q^0-1-n_0^{\rm e}\beta v_{\rm ee}(Q)G_{\rm x}(Q)\chi_Q \,.
\end{equation}
Here, $G_{\rm x}(Q)$ is the exchange part of the local field correction.
If we approximate $G_{\rm x}(Q)$ by the use of $G_{\rm x}^{\rm jell}(Q)$, which is well known for an electron gas in the jellium model, the QHNC method yields a closed set of equations for plasma properties. To get a closed set of equations to determine all quantities in a self-consistent manner it is necessary to build up a new equation for $G_{\rm x}(Q)$. 

%
\section{Charge neutrality condition in the electron-ion model}\label{s:Sboundfree}
%
At first, we note exact relations between the structure factors $S_{ij}(Q)$ in the electron-ion model \cite{QHNC}: 
\begin{eqnarray}
S_{\rm eI}(Q)&=&{ \rho(Q) \over \sqrt{Z_{\rm I}} } S_{\rm II}(Q) \label{e:sei2} \\
\chi_{\rm ee}(Q)&=& {|\rho(Q)|^2 \over Z_{\rm I} }S_{\rm II}(Q)+{\chi_Q^0 \over 1-n_0^{\rm e} C_{ee}(Q)\chi_Q^0}\,.   \label{e:xee}
\end{eqnarray}
Here, $\rho(Q)$ is the screening density distribution of a pseudo-atom defined by the non-interacting density response function $\chi_Q^0$ and the direct correlation functions $C_{ij}$ in the electron-ion mixture:
\begin{equation}
\rho(Q)\equiv { n_0^{\rm e} C_{\rm eI}(Q)\chi_Q^0 \over 1-n_0^{\rm e} C_{\rm ee}(Q)\chi_Q^0 } \,.
\end{equation}
Thus, Eq.~(\ref{e:sei2}) leads to the exact relation, which must be followed in any electron-ion model:
\begin{equation}\label{e:ChNeuS}
Z_{\rm I}S_{\rm II}(0)=\sqrt{Z_{\rm I}}S_{\rm eI}(0)=S_{\rm ee}(0)=\chi_{\rm ee}(0)=n_0^{\rm e}\kappa_{\rm T}/\beta \,,
\end{equation}
with the compressibility $\kappa_{\rm T}$.
By the inverse Fourier transform, 
the first part of the above equation is rewritten in the form
\begin{equation}\label{e:chargeN}
Z_{\rm I}=n_0^{\rm e}\int [g_{\rm eI}(r)-1]d{\bf r}-Z_{\rm I}n_0^{\rm I}\int 
[g_{\rm II}(r)-1]d{\bf r} \,,
\end{equation}
which states that an ion fixed at the origin keeps the charge 
neutrality by accumulating the free electrons and by pushing 
away the ions around it in the whole space, not within the 
Wigner-Seitz cell.

On the other hand, 
when we fix an ion in the electron-ion mixure at the origin of coordinates, the nuclei forming this ion has the electron density  $n_{\rm e}(r)$ around it;
this electron density $n_{\rm e}(r)$ is obtained by solving a wave equation for the external potential caused by this fixed nucleus as a sum of the bound electron density $n_{\rm e}^{\rm b}(r)$ and the continuum electron density $n_{\rm e}^{\rm c}(r)$. Furthermore, the electron density  $n_{\rm e}(r)$ satisfies the following equation represented in terms of the Friedel sum of phase shifts $\delta_{\ell}(E)$ 
\begin{eqnarray}
\int [n_{\rm e}(r)\!-\!n_0^{\rm e}]d{\bf 
r}&=&\!\!\!\sum_{\epsilon_i<0}\!f(\epsilon_i)+\frac2{\pi}\sum_{\ell}
(2\ell\!+\!1)\,\!
\int_0^{\infty}\!\!\!f(E){d\delta_{\ell}(E)\over dE}dE \,,
\end{eqnarray}
where $n_{\rm e}(r)\!=\!n_{\rm e}^{\rm b}(r)\!+\!n_{\rm e}^{\rm c}(r)\!=\!\rho_{\rm b}(r)\!+\!n_{\rm e}^{\rm f}(r|{\rm N})$ is a sum of the "bound electron" distribution $\rho_{\rm b}(r)$ and the free-electron distribution $n_{\rm e}^{\rm f}(r|{\rm N})$ with  $f(\epsilon)\!=\!1/[\exp\{\beta (\epsilon\!-\!\mu_{\rm e}^0)\}\!+\!1]$.
If we take the bound-electron density $n_{\rm e}^{\rm b}(r)$ to define $Z_{\rm B}\!\!=\!\!\int n_{\rm e}^{\rm b}(r)d{\bf r}\!\!=\!\!\sum_{\epsilon_i<0}\!f(\epsilon_i)$, the free-electron part $n_{\rm e}^{\rm f}(r)\!=\!n_{\rm e}^{\rm c}(r)$ must satisfy the following relation:
\begin{equation}
Z_{\rm I}S_{\rm II}(0)=\int [n_{\rm e}^{\rm f}(r)-n_0^{\rm e}]d{\bf r}=Z_{\rm I}n_0^{\rm I}\kappa_{\rm 
T}/\beta=\frac2{\pi}\sum_{\ell}(2\ell+1)\!
\int_0^{\infty}\!\!\!f(E){d\delta_{\ell}(E)\over dE}dE \,.
\label{e:fs} 
\end{equation}
This relation is fulfilled generally for simple metals. However, there are some liquid metals and plasmas, 
for which this relation cannot be satisfied due to the large contribution of resonant phase-shifts and small compressibility $\kappa_{\rm T}$ at the high density.
As a consequence, we must treat in general a part $\Delta\rho_{\rm b}(r)$ of the continuum electron $n_{\rm e}^{\rm c}(r)$ as the part involved in the bound-electron distribution $\rho_{\rm b}(r)$ to form an ion: $\rho_{\rm b}(r)=n_{\rm e}^{\rm b}(r)\!\!+\!\!\Delta\rho_{\rm b}(r)$ and $Z_{\rm B}\!\!=\!\!\int [n_{\rm e}^{\rm b}(r)\!\!
+\!\!\Delta\rho_{\rm b}(r)]d{\bf r}$.

For the purpose of obtaining the expression of $\Delta\rho_{\rm b}(r)$, let us consider 
a chosen central ion as an atom immersed in the electron-ion mixture with a nucleus $Z_{\rm A}$ fixed at the origin of coordinates: the electron-ion model with a nucleus, forming the central ion, fixed at the origin is referred to as "the average atom (AA) model", hereafter.  
This nucleus accumulates electrons with the electron density $n_{\rm e}(r)\!\!\equiv \!\!n_{\rm e}(r|{\rm N})$ and pushes away surrounding ions with $n_{\rm I}(r|{\rm N})$, keeping the charge neutrality condition around it: 
\begin{equation}
Z_{\rm A}=\int [ n_{\rm e}(r|{\rm N})-n_0^{\rm e}]d{\bf r}-Z_{\rm I}\int [n_{\rm I}(r|{\rm N})-n_0^{\rm I}]d{\bf r}\,. \label{e:neutNuc} 
\end{equation}
The following three conditions are necessary for the electron density $n_{\rm e}(r|{\rm N})$ to be consistent with the charge neutrality condition (\ref{e:chargeN}) in the electron-ion model.

\noindent (I): As is seen from (\ref{e:neutNuc}), the central nucleus forming an ion around it produces an external potential $v_{\rm IN}(r)$ to the ions, and pushes away surrounding ions in the same manner to (\ref{e:chargeN}) of the electron-ion model, where the central ion yielding an external potential $v_{\rm II}(r)$ pushes away the other ions. This means that the external potential $v_{\rm IN}(r)$ for the ions caused by this fixed nucleus should be identical to the ion-ion interaction $v_{\rm II}(r)$ as an external potential to the other ions.
In a plasma which can be regarded as an electron-ion mixture, the ion distribution $n_{\rm I}(r|\rm I)$ around a fixed ion is identical with the ion distribution $n_{\rm I}(r|\rm N)$ around the fixed nucleus forming this fixed ion, since we see the same ion-distribution only from two different points of view: we sit either on the ion or on the nucleus inside the ion. Therefore, we have $n_{\rm I}(r|{\rm N})\!\equiv\!n_0^{\rm I}g_{\rm IN}(r)\!=\!n_0^{\rm I}g_{\rm II}(r)\!\equiv\!n_{\rm I}(r|\rm I)$. As a natural result, we have the relation: $v_{\rm IN}(r)\!=\!v_{\rm II}(r)$ (see Appendix A).
\\
(II): From the same reason above, in addition to  $v_{\rm eN}(r)\!=\!v_{\rm eI}(r)$ the free-electron distribution $n_{\rm e}^{\rm f}(r|{\rm N})$ around a fixed nucleus is identical to the electron-ion RDF $n_0^{\rm e}g_{\rm eI}(r)$: $n_{\rm e}^{\rm f}(r|{\rm N})\!=\!n_{\rm e}(r|{\rm N})\!-\!\rho_{\rm b}(r)\!=\!n_0^{\rm e}g_{\rm eI}(r)$\,.
 Thus, Eqs.~(\ref{e:neutNuc}) and (\ref{e:chargeN}) lead to the definition of the number of bound electron in the ion, $Z_{\rm B}\!=\!\int \rho_{\rm b}(r)d{\bf r}$, in the form:
\begin{equation}\label{e:Zb}
Z_{\rm B}= \int [n_{\rm e}(r)\!-\!n_0^{\rm e}g_{\rm eI}(r)]d{\bf r} \,,
\end{equation}
with the bound-electron distribution $\rho_{\rm b}(r)$ to yield the ion
\begin{equation}\label{e:rho-b}
\rho_{\rm b}(r) = n_{\rm e}(r)\!-\!n_0^{\rm e}g_{\rm eI}(r) \,.
\end{equation}
(III): $n_0^{\rm e}\!=\![Z_{\rm A}\!-\!Z_{\rm B}]n_0^{\rm I}\!=\!Z_{\rm I}n_0^{\rm I}$.
 As a result, Eq.~(\ref{e:neutNuc}) can be rewritten in the form:
\begin{equation}\label{e:Za}
Z_{\rm A}= \int [n_{\rm e}(r)\!-\!n_0^{\rm e}g_{\rm II}(r)]d{\bf r}\,.
\end{equation}

As is shown above, the average ion charge $Z_{\rm I}$ is given in terms of $g_{\rm eI}(r)$ [a quantity of the electron-ion model] and $Z_{\rm A}$ with $n_{\rm e}(r)$ [quantities of the AA model] as follow:
\begin{equation}\label{e:DfZi}
Z_{\rm I}=\frac{n_0^{\rm e}}{n_0^{\rm I}}=Z_{\rm A}-\int [n_{\rm e}(r)\!-\!n_0^{\rm e}g_{\rm eI}(r)]d{\bf r}=Z_{\rm A}-\int \rho_{\rm b}(r)d{\bf r} \,.
\end{equation}
This relation plays an important role to correlate the AA model to the electron-ion model. It should be noted that this relation comes only from (\ref{e:chargeN}), which is an exact relation for any electron-ion mixture.
On the other hand, the free-electron distribution $n_{\rm e}^{\rm f}(r)$ around a fixed nucleus is defined as 
\begin{equation}\label{e:nef}
n_{\rm e}^{\rm f}(r)\equiv n_{\rm e}^{\rm c}(r)\!\!-\!\!\Delta\rho_{\rm b}(r)\!=\!
n_{\rm e}(r)\!\!-\!\!n_{\rm e}^{\rm b}(r)\!\!-\!\!\Delta\rho_{\rm b}(r)\!=\!n_0^{\rm e}g_{\rm eI}(r) \,.
\end{equation}
Hence, $\Delta\rho_{\rm b}(r)$ is represented in terms of quantities of the AA model [$n_{\rm e}(r)$ and $n_{\rm e}^{\rm b}(r)$] and quantities of the electron-ion model [$\rho(Q)$ and $S_{\rm II}(Q)$] as follows:
\begin{equation}
\Delta\rho_{\rm b}(r)\equiv n_{\rm e}(r)\!\!-\!\!n_{\rm e}^{\rm b}(r)-n_0^{\rm e}\!-\!{\cal F}_{\bf r}[\,\rho(Q)S_{\rm II}(Q)\,] \,.
\end{equation}
Here, the chemical potential $\mu_{\rm e}^0$ satisfies the following equation:
\begin{eqnarray}
Z_{\rm A}=&&\sum_{\epsilon_i<0} {1\over \exp[\beta 
(\epsilon_i-\mu_{\rm e}^0)]+1} +\Delta Z_{\rm B}  \nonumber \\&&+\frac1{n_0^{\rm I}}\int {2\over 
\exp[\beta (p^2/2m-\mu_{\rm e}^0)]+1} 
\frac{d{\bf p}}{(2\pi\hbar)^3}\,, \label{e:gcp}
\end{eqnarray}
since $\Delta Z_{\rm B}\equiv \int \Delta\rho_{\rm b}(r) d{\bf r}$ comes from the resonance in the 
continuum [(3.87) in \cite{QHNC}], or from a part of the bound electrons $n_{\rm e}^{\rm b}(r)$, which does not contribute to the formation of an ion [Appendix~B]. For example, in a fully ionized hydrogen plasma, the bound-electron distribution $\rho_{\rm b}(r)$ is zero even though $n_{\rm e}^{\rm b}(r)\!\neq\! 0$, since the electron-proton RDF is given by $n_0^{\rm e}g_{\rm ep}(r)\!=\!n_{\rm e}^{\rm b}(r)\!+\!n_{\rm e}^{\rm c}(r) $ \cite{ChiharaHplasma}.
In a simple metal on the other hand, there follows $\Delta\rho_{\rm b}(r)\!=\!0$, because of $Z_{\rm B}\!\!=\!\!\int n_{\rm e}^{\rm b}(r)d{\bf r}$. Another example is given for the case of a high density plasma which has a strong resonant state and a small compressibility $\kappa_{\rm T}\!\approx\!0$: for this system we obtain the bound-electron number $Z_{\rm B}$, as is seen from Eq.~(\ref{e:apprChargeN}) below
\begin{eqnarray}
Z_{\rm B}\!\!&\!=&\!\!\!\!\int [n_{\rm e}(r)\!-\!n_0^{\rm e}]d{\bf r}
=\!\sum_{\epsilon_i<0}\!f(\epsilon_i)\!+\!\frac2{\pi}\sum_{\ell}
(2\ell\!+\!1)\,\!\int_0^{\infty}\!\!\!f(E){d\delta_{\ell}(E)\over dE}dE \label{e:ZbBlens} \\
&\approx&\sum_{\epsilon_i<0}\!f(\epsilon_i)\!+\!\frac2{\pi}
(2\ell_0\!+\!1)\,\!\int_0^{\infty}\!\!\!f(E){d\delta_{\ell_0}(E)\over dE}dE \,.
\end{eqnarray}

In final, note that Eq.~(\ref{e:DfZi}) is written as:
\begin{equation}\label{e:ChargeN2}
Z_{\rm I}=\frac{n_0^{\rm e}}{n_0^{\rm I}}=Z_{\rm A}-\int [n_{\rm e}(r)\!-\!n_0^{\rm e}]d{\bf r}
+Z_{\rm I}S_{\rm II}(0) \,.
\end{equation}
When $S_{\rm II}(0)\!=\!n_0^{\rm I}\kappa_{\rm T}/\beta\!\approx\! 0$, 
Eq.(\ref{e:ChargeN2}) becomes 
\begin{equation}\label{e:apprChargeN}
Z_{\rm I}=\frac{n_0^{\rm e}}{n_0^{\rm I}}=Z_{\rm A}-\int [n_{\rm e}(r)\!-\!n_0^{\rm e}]d{\bf r}\,,
\end{equation}
which is adopted by Blenski and coworkers \cite{Blenski07a,Blenski07b,Piron2011} as the charge neutral condition in the VAAQP method.
Although this equation has the same structure to (\ref{e:DfZi}), it does not mean $\rho_{\rm b}(r)\!=\!n_{\rm e}(r)\!-\!n_0^{\rm e}$\,.
%
\section{Average ion in the jellium-vacancy model}\label{JVion}
%

On the base of a charge neutrality condition (\ref{e:apprChargeN}) and the local density approximation (LDA), Blenski and coworkers \cite{Blenski07a,Blenski07b,Piron2011} derived a formula to determine the uniform electron density $n_0(=\!n_0^{\rm e})$ by use of variational principles applied to the free energy of a plasma as a mixture of electrons and ions, with respect to $n_0$. 
Here, we show that the formula determining the electron density $n_0$ can be derived exactly within the framework of DF theory \cite{chiharaDF} and the jellium-vacancy model by following the procedure of Blenski {\it et al}, without use of LDA. Their formula is obtained by appling two approximations to our formula: hence, the application limit of their formula becomes clear.

In a neutral electron-ion mixture, the electrons in this mixture can be treated as an electron gas in the uniform positive background (jellium model) as an approximation.
Let us consider an electron gas in the uniform background which has a vacancy  $g(r)$ [the step-function $\theta (r\!-\!R)$, for example] with a fixed nucleus $Z_{\rm A}$ at the center of the vacancy.
In this system, the central nucleus accumulates surrounding electrons to form an "average ion"  in a plasma: this is referred to as the jellium-vacancy model (or the neutral pseudo-atom model). The electron density distribution $n_{\rm e}(r)$ around the central nucleus is produced by imposing an external potential $U(r)$ to a electron gas in the uniform background: here $U(r)$ comes from the vacancy with a nucleus fixed at its center.
Therefore, we can determine the electron density distribution $n_{\rm e}(r|{U})$ under the external potential ${U}$ by the DF theory using its free energy $F$\,.

In treating this inhomogeneous system, an effective external potential $U_{\rm eff}({\bf r})$ can be defined so as to satisfy 
the condition that the true electron density $n_{\rm e}({\bf r}|U)$ under the 
external potential $U({\bf r})$ should be identical with 
the non-interacting electron density distribution $n^0({\bf r}|U_{\rm eff})$ under the external potential $U_{\rm eff}({\bf r})$:
\begin{equation}\label{e:dnsn0n}
n^0({\bf r}|U_{\rm eff})\equiv n_{\rm e}({\bf r}|U)\,.
\end{equation}
Here, the non-interacting electron density distribution 
$n^0({\bf r}|U_{\rm eff})$ is determined as  
\begin{equation}\label{e:n0Ueff}
n^0({\bf r}|U_{\rm eff})\equiv \sum_i f_i|\phi_i({\bf r})|^2\,,
\end{equation}
in terms of the wave function $\phi_i({\bf r})$ and eigenvalues $\epsilon_i$, 
which obey the wave equation for a single electron:
\begin{equation}\label{e:KSwe}
\left[\frac{-\hbar^2}{2m}\nabla^2+U_{\rm eff}({\bf r})\right]\phi_i({\bf r})=\epsilon_i\phi_i({\bf r})\,,
\end{equation}
and the Fermi distribution $f_i\!\!=\!\!f(\epsilon_i)$ 
with the chemical potential $\mu_0$ of a non-interacting electron gas.

For the purpose to obtain an explicit expression for the external potential 
$U_{\rm eff}({\bf r})$, let us note the relation for intrinsic free energy 
of non-interacting electrons ${\cal F}_s$:
\begin{equation}\label{e:gamm0}
\left.\frac{\delta{\cal F}_s}{\delta n^0({\bf r})}\right|_{TV}
=\left.\frac{\delta{\cal F}_s}{\delta n_{\rm e}({\bf r})}\right|_{TV}
=\mu_0-U_{\rm eff}({\bf r})\equiv\gamma_{\rm eff}({\bf r})\,,
\end{equation}
which is the rewritten form for the non-interacting system  of Eq.~(\ref{e:gamm}), below, in the real system [(3.5) in \cite{chiharaDF}].
Here, the intrinsic free energy ${\cal F}_s[n^0]$ of the non-interacting system is written in an explicit form:
\begin{eqnarray}
{\cal F}_s[n^0]&=&T_s[n^0]-TS_s[n^0]\label{e:intFs2}\,,
\end{eqnarray}
where $T_s$ and $S_s$ are the intrinsic internal energy $\tilde E_s$ and the entropy of the non-interacting system, respectively, defined by 
\begin{eqnarray}
T_s[n^0]&\equiv&{\tilde E}_s \equiv E_s
-\int n^0({\bf r})U_{\rm eff}({\bf r})d{\bf r}\label{e:keTs}\\
&=&\sum_if_i\int \phi^*_i({\bf r})\frac{-\hbar^2}{2m}\nabla^2\phi_i({\bf r})d{\bf r}\,,\\
S_s[n^0]&\equiv& -k_{\rm B}\sum_i[f_i\ln f_i+(1-f_i)\ln (1-f_i)]\,\label{e:entSs}
\end{eqnarray}
with the internal energy $E_s=\sum_if_i\epsilon_i$.
On the other hand, the intrinsic free energy ${\cal F}\equiv {F}-\int U({\bf r})n_{\rm e}({\bf r})d{\bf r}$ defined as a part of the free energy $F$ of this system leads to 
\begin{equation}\label{e:gamm}
\left.\frac{\delta{\cal F}}{\delta n_{\rm e}({\bf r})}\right|_{TV}
=\mu[n(r|U)]-U({\bf r})\equiv\gamma({\bf r})\,,
\end{equation}
as is shown by (2.13) in Reference \cite{{chiharaDF}}. Thus, the effective external potential $U_{\rm eff}({\bf r})$ is represented explicitly as
\begin{equation}\label{e:Ueff1}
U_{\rm eff}({\bf r})=U({\bf r})+\left.\frac{\delta {\cal F}_{\rm I}}{\delta n_{\rm e}({\bf r})}\right|_{TV}-\mu_{\rm I} 
\end{equation}
with ${\cal F}_{\rm I}\!\!\equiv\!\!{\cal F}\!-\!{\cal F}_s$ and $\mu_{\rm I}\!\!\equiv\!\!\mu[n(r|U)]\!-\!\mu_0$, which is derived by subtracting (\ref{e:gamm0}) from (\ref{e:gamm}).
Equation (\ref{e:Ueff1}) is the effective external potential to produce the real electron density $n_{\rm e}({\bf r}|U)\!=\!n^0({\bf r}|U_{\rm eff})$ in terms of the non-interacting electron density distribution.
At this point, we introduce the exchange-correlation free energy: 
\begin{equation}\label{e:tFxc}
{\cal F}_{\rm xc}\equiv {\cal F}_{\rm I}-\tilde E_{\rm es}\,,
\end{equation}
in order to extract the exchange-correlation effect from the interaction part ${\cal F}_{\rm I}$ by subtracting the intrinsic electrostatic energy $\tilde E_{\rm es}\!\equiv\! E_{\rm es}[n]\!-\!\int U({\bf r})n_{\rm e}({\bf r})d{\bf r}$ defined for the electrostatic energy $E_{\rm es}[n]$ \cite{chiharaDF}.
Then, we can represent the intrinsic free energy ${\cal F}$ in the final form:
\begin{equation}\label{e:tFdf}
{\cal F} \equiv {F}\!-\!\int U({\bf r})n_{\rm e}({\bf r})={\cal F}_s+{\cal F}_{\rm I}={\cal F}_s+{\cal F}_{\rm xc}+\tilde E_{\rm es}\,,
\end{equation}
by taking the non-interacting particles, ${\cal F}_s$, as a reference system in the DF theory.
Thus, there results from (\ref{e:Ueff1}) [(3.30) in \cite{chiharaDF}]:
\begin{equation}\label{e:Ueff2}
U_{\rm eff}({\bf r})=\left.\frac{\delta E_{\rm es}}{\delta n_{\rm e}({\bf r})}\right|_{U({\bf r})}+\left.\frac{\delta {\cal F}_{\rm xc}}{\delta n_{\rm e}({\bf r})}\right|_{TV}-\mu_{\rm I}=\left.\frac{\delta [E_{\rm es}+{\cal F}_{\rm xc}]}{\delta n_{\rm e}({\bf r})}\right|_{TVU(r)}-\mu_{\rm I}\,.
\end{equation}

In the above, we described general thermodynamic relations in the DF theory as shown in the reference \cite{chiharaDF}. At this point, we consider an electron gas in the jellium 
with a vacancy $g(r)$ fixed a nucleus ${Z_{\rm A}}$ at the center of the vacancy: $g(r)$ must be chosen to satisfy a condition $Z_{\rm A}\!=\!\int [n_{\rm e}(r)\!-\!n_0 g(r)]d{\bf r}$.
Since the electrostatic energy $E_{\rm es}$ for this system is given by
\begin{equation}\label{e:Ees}
E_{\rm es}=\int\! d{\bf r}\Biggl\{ \left(n_{\rm e}(r)\!-\!n_0g(r)\right)\Biggr.
\Biggl.\left( -\frac{Z_{\rm A}}{r}+\frac{1}{2}\int\,d{\bf r}'v_{\rm ee}(|{\bf r}\!-\!{\bf r'}|)\left\{ {n_{\rm e}(r')\!-\!n_0 g(r')} \right\} \right) \Biggr\}\,,
\end{equation}
the electrostatic part of the effective potential (\ref{e:Ueff2}) is written as
\begin{eqnarray}
\left.\frac{\delta E_{\rm es}}{\delta n_{\rm e}({\bf r})}\right|_{U({\bf r})}
&=& -\frac{Z_{\rm A}}{r}+\int v_{\rm ee}(|{\bf r}\!-\!{\bf r'}|)[n_{\rm e}(r')\!-\!n_0 g(r')]d{\bf r}'\\
&=& U(r)+\int v_{\rm ee}(|{\bf r}\!-\!{\bf r'}|)[n_{\rm e}(r')\!-\!n_0 ]d{\bf r}' \,. \label{e:dEesDdn} 
\end{eqnarray}
Here, 
\begin{equation}
U(r) \equiv -Z_{\rm A}/r+f(r)
\end{equation}
\begin{equation}
f(r)\equiv -\!n_0\!\!\int \!v_{\rm ee}(|{\bf r}\!-\!{\bf r'}|)[g(r')\!-\!1]d{\bf r}' \,.
\end{equation}
That is, a vacancy $g(r)$ with a nucleus $Z_{\rm A}$ fixed at the center of the vacancy causes an external potential $U(r)$ to the electron gas in the uniform background. 
As a consequence, we can obtain the electron density $n_{\rm e}(r|{\rm U})$ by solving a wave equation for non-interacting electrons under the external potential  $U_{\rm eff}(r)$:
\begin{equation}\label{e:Ueff3}
U_{\rm eff}({\bf r})=U(r)+\int v_{\rm ee}(|{\bf r}\!-\!{\bf r'}|)[n_{\rm e}(r')\!-\!n_0 ]d{\bf r}'+\mu_{\rm xc}({\bf r}|n_{\rm e})-\mu_{\rm I} \,.
\end{equation}
Here, $\mu_{\rm xc}({\bf r}|n_{\rm e})\!\!\equiv\!\! {\delta {\cal F}_{\rm xc}}/{\delta n_{\rm e}({\bf r})}|_{TV}$ is the exchange-correlation potential. If we choose the chemical potential $\mu$ so as to be $\lim_{r\rightarrow\infty}U_{\rm eff}({\bf r})\!=\!0$, there result $\mu_{\rm I}\!=\!\mu_{\rm xc}[n_0]$ and $\mu[n(r|U)]\!=\!\mu[n_0]\!\equiv\! \mu$ because of $\lim_{r\rightarrow\infty}\!U(r)\!=\!0$.

When a nucleus immersed in an electron gas in the jellium creates an ion, the free energy $F[n_0]$ of the uniform electron gas with the density $n_0$ becomes $F[n_{\rm e}(r)]$;  the change of the free energy due to the formation of one ion in a plasma is given by $F[n_{\rm e}(r)]\!-\!F[n_0]$. 
Therefore, the free energy of a plasma with $N$ nuclei is given by $A[n_{\rm e}(r),n_0]\!\equiv\! F[n_0]/N\!+\!\{ F[n_{\rm e}(r)]\!-\!F[n_0] \}$ per particle, since all ions in a plasma have the same structure (the cluster expansion, \cite{Blenski07b}). At this point, we derive an equation for $n_0$ by minimizing the free energy $A[n_{\rm e}(r),n_0]$ with respect to $n_0$, under the condition to fulfill the charge neutrality condition (\ref{e:Za}). For this purpose, we introduce the following function with the Lagrange multiplier $\gamma$:
\begin{equation}\label{e:nc}
\Omega[n_{\rm e}(r),n_0]\equiv F[n_0]/N+\{ F[n_{\rm e}(r)]\!-\!F[n_0] \}-\gamma \left( \left[\int [n_{\rm e}(r)\!-\!n_0 g(r)]d{\bf r}-Z_{\rm A}\right] \right) \,.
\end{equation}
Here, $F\!=\!{\cal F}\!+\!\int U({\bf r})n_{\rm e}({\bf r})d{\bf r}\!=\!{\cal F}_s[n_{\rm e}(r)]\!+\!{\cal F}_{\rm xc}[n_{\rm e}(r)]\!+\!E_{\rm es}$ due to (\ref{e:tFdf}).

In general, for $A[n_{\rm e}(r),n_0]$, which is a functional of $n_{\rm e}(r)$ and also a function of $n_0$, we have the next relation:
\begin{equation}\label{e:varNn0}
\frac{\delta A[n_{\rm e}(r),n_0]}{\delta n_0}=\int \frac{\delta A[n_{\rm e}(r),n_0]}{\delta n_{\rm e}({\bf r}')}\frac{\delta n_{\rm e}({\bf r}')}{\delta n_0}d{\bf r}'+\frac{\partial A[n_{\rm e}(r),n_0]} {\partial n_0} \,.
\end{equation}
So, Eq.(\ref{e:varNn0}) leads to the variation of $E_{\rm es}$ given by (\ref{e:Ees}) with respect to $n_0$ as:
\begin{equation}
\left.\frac{\delta E_{\rm es}}{\delta n_0}\right|_{U({\bf r})}=\int \left.\frac{\delta E_{\rm es}}{\delta n_{\rm e}({\bf r}')}\right|_{U({\bf r})}\frac{\delta n_{\rm e}({\bf r}')}{\delta n_0}d{\bf r}'-\int \left.\frac{\delta E_{\rm es}}{\delta n_{\rm e}({\bf r})}\right|_{U(\bf r)}\!\!g(r)d{\bf r} \,.
\end{equation}
Here, 
\begin{equation}\label{e:Ves}
\left.\frac{\delta E_{\rm es}}{\delta n_{\rm e}({\bf r})}\right|_{U(\bf r)}
=-\frac{Z_{\rm A}}{r}+f(r)+\int v_{\rm ee}(|{\bf r}\!-\!{\bf r'}|)[n_{\rm e}(r)-n_0]d{\bf r}'
\equiv v_{\rm es}(r)\,.
\end{equation}
Next, the variation of ${\cal F}_s[n_{\rm e}(r)]\!+\!{\cal F}_{\rm xc}[n_{\rm e}(r)]$ with respect to $n_0$ is obtained only through the variation of $n_{\rm e}(r)$. Therefore, the result is written as
\begin{eqnarray}
\left.\frac{\delta F}{\delta n_0}\right|_{TVU({\bf r})}\!\!\!\!\!\!&=&\!\!\int\!\! \left.\frac{\delta [{\cal F}_s+{\cal F}_{\rm xc}]}{\delta n_{\rm e}({\bf r}')}\right|_{TVU({\bf r})}\!\!\!\!\frac{\delta n_{\rm e}({\bf r}')}{\delta n_0}d{\bf r}'+\int \!\!\left.\frac{\delta E_{\rm es}}{\delta n_{\rm e}({\bf r}')}\right|_{U({\bf r})}\!\!\!\!\frac{\delta n_{\rm e}({\bf r}')}{\delta n_0}d{\bf r}'-\int \left.\frac{\delta E_{\rm es}}{\delta n_{\rm e}({\bf r})}\right|_{U(\bf r)}\!\!g(r)d{\bf r} \nonumber\\
&=& \int \left.\frac{\delta {F}}{\delta n_{\rm e}({\bf r}')}\right|_{TVU({\bf r})}\frac{\delta n_{\rm e}({\bf r}')}{\delta n_0}d{\bf r}'-\int \left.\frac{\delta E_{\rm es}}{\delta n_{\rm e}({\bf r})}\right|_{U(\bf r)}\!\!g(r)d{\bf r}\\
&=& \mu[n(r|U)] \int \frac{\delta n_{\rm e}({\bf r}')}{\delta n_0}d{\bf r}'
 - \int \left.\frac{\delta E_{\rm es}}{\delta n_{\rm e}({\bf r})}\right|_{U(\bf r)}\!\!g(r)d{\bf r} \,.
\end{eqnarray}
In the above derivation we use the next equation [(2.14) in Reference \cite{chiharaDF}]:
\begin{equation}\label{e:muU}
\left.\frac{\delta F}{\delta n_{\rm e}({\bf r})}\right|_{TVU({\bf r})}=\mu[n(r|U)]=\left.\frac{\delta{\cal F}}{\delta n_{\rm e}({\bf r})}\right|_{TV}+U({\bf r})\,.
\end{equation}
On the other hand, the variation of the electron-gas free energy $F[n_0]$ with respect to $n_0$ is written in the form:
\begin{equation}
\left.\frac{\delta F[n_0]}{\delta n_0}\right|_{TV} 
 = \int \left.\frac{\delta F[n_{\rm e}]}{\delta n_{\rm e}({\bf r})}\right|_{TV,U=0}d{\bf r}
=\left.\frac{\partial F[n_0]}{\partial n_0}\right|_{TV}= \int \mu d{\bf r}\,.
\end{equation}
In final, summing up all the above results, the variation of $\Omega$ about $n_0$ is obtained as
\begin{eqnarray}
\left.\frac{\delta \Omega}{\delta n_0}\right|_{TVU({\bf r})} 
&=& \frac{\int \mu d{\bf r}}{N}+\mu[n(r|U)] \int \frac{\delta n_{\rm e}({\bf r}')}{\delta n_0}d{\bf r}'
 - \int \left.\frac{\delta E_{\rm es}}{\delta n_{\rm e}({\bf r})}\right|_{U(\bf r)}\!\!g(r)d{\bf r}-\int \mu d{\bf r} \nonumber \\
&&-\gamma \int \left[\frac{\delta n_{\rm e}({\bf r}')}{\delta n_0}-1\right]d{\bf r}'-\frac{\gamma [1-S(0)]}{n_0^{\rm I}}=0 \,.
\end{eqnarray}
Since $\mu[n(r|U)]\!=\!\mu$ for the case $\lim_{r\rightarrow\infty}\!U(r)\!=\!0$, the above equation provides the final equation in conjunction with $\gamma\!=\!\mu$ :
\begin{equation}\label{e:cnc1}
\int \left.\frac{\delta E_{\rm es}}{\delta n_{\rm e}({\bf r})}\right|_{U(\bf r)}\!\!g(r)d{\bf r}=\int v_{\rm es}(r)g(r)d{\bf r}=\mu S(0)/n_0^{\rm I}\,.
\end{equation}
Here,
\begin{equation}
S(Q)\equiv 1+n_0^{\rm I}{\cal F}_{\bf Q}[g(r)-1]\,.
\end{equation}
Therefore, when the vacancy function $g(r)$ is determined in a self-consistent way in the form, $g(r)\!=\!g_{\rm II}(r)$, the charge neutrality condition (\ref{e:Za}) is exactly satisfied.
Also, when $g(r)$ becomes the RDF $g_{\rm II}(r)$ in an electron-ion mixture, the uniform electron density $n_0$ must fulfill the following equation:
\begin{equation}\label{e:cndExact}
\int v_{\rm es}(r)g_{\rm II}(r)d{\bf r}=\mu S_{\rm II}(0)/n_0^{\rm I}=\mu \kappa_{\rm T}/\beta \,.
\end{equation}
It should be mentioned that Eq.~(\ref{e:cndExact}) is derived without any approximation to ${\cal F}_{\rm xc}[n_{\rm e}(r)]$ (the  exchange-correlation part of free energy), and furthermore it is derived only from the structure of electronic state $E_{\rm es}$ (\ref{e:Ees}) and the charge neutrality condition (\ref{e:Za}). This fact can be seen directly: when the electronic energy is given by $E_{\rm es}\!=\!E_{\rm es}[n_{\rm e}(r),n_0]$ and the charge neutrality condition is $C[n_{\rm e}(r),n_0]\!=\!0$, the condition to determine the uniform electron density $n_0$  is written as
\begin{equation}\label{e:n0C}
\frac{ \partial E_{\rm es}[n_{\rm e}(r),n_0] }{\partial n_0}=\mu \left [\frac{\partial C[n_{\rm e}(r),n_0]}{\partial n_0} + \int d{\bf r}-\frac{1}{ n_0^{\rm I}}   \right] \,.
\end{equation}
When we adopt (\ref{e:DfZi}) instead of (\ref{e:Za}) for the charge neutrality condition $C[n_{\rm e}(r),n_0]\!=\!0$, Eq.~(\ref{e:n0C}) yields the next equation:
\begin{equation}\label{e:NEUgei}
\int \left.\frac{\delta E_{\rm es}}{\delta n_{\rm e}({\bf r})}\right|_{U(\bf r)}\!\!g_{\rm II}(r)d{\bf r}=\int v_{\rm es}(r)g_{\rm II}(r)d{\bf r}=\mu \int [g_{\rm eI}(r)-1]d{\bf r}/n_0^{\rm I}\,.
\end{equation}
Furthermore, it is shown that the above equation reduces to (\ref{e:cndExact}) with the use of (\ref{e:ChNeuS}). In another point of view, it is necessary that the vacancy $g(r)$ in (\ref{e:Ees}) should be identical with the RDF $g_{\rm II}(r)$ for the neutrality condition (\ref{e:NEUgei}) to be the same as the other (\ref{e:cndExact}).
As shown above, Eq.~(\ref{e:cndExact}) is exact in the framework of the jellium-vacancy model and the DF theory based on the jellium model. In contrast with our derivation, Blenski and Cichocki \cite{Blenski07b} derived Eq.~(\ref{e:cndExact}) on the base of (\ref{e:Ueff2}) and (\ref{e:gamm0}) using some approximations.

Equation (\ref{e:cndExact}) reduces to the result of Blenski and coworkers \cite{Blenski07a,Blenski07b,Piron2011}, when we apply the following two approximations to (\ref{e:cndExact}): (A) $g_{\rm II}(r)\!=\!\theta(r\!-\!R_{\rm ws})$ and (B) $S_{\rm II}(0)\!=\!n_0^{\rm I}\kappa_{\rm T}/\beta\!=\!0$. Note that the approximation (B) is a good one for a plasma at high density and low temperature because of $\kappa_{\rm T}/\beta\!\approx\!0$, but at this state the approximation (A) becomes inappropriate in the integration of the left-hand side of (\ref{e:cndExact}). This is the reason why the VAAQP method does not provide a good description of a plasma at high density and low temperature. The approximation (A) yields a large error in the evaluation of  the electrostatic energy (\ref{e:Ees}). In order to treat a plasma at high density and low temperature, the RDF $g_{\rm II}(r)$ must be correctly determined. Moreover, the 'ionization model' (\ref{e:apprChargeN}) cannot treat the ionization caused by the temperature effect, as is seen from (\ref{e:ZbBlens}).

%
\section{Average atom in the electron-ion model and QHNC method}\label{s:QHNC}
%
In this section we give a summary to set up a set of integral equations (QHNC) for calculating the RDFs in the electron-ion model with some improvement.
At first, let us determine the density distribution $n_i(r|U_{\rm I}U_{\rm e})$ of $i$-species particle ($i$=e or I) induced by external potentials, $U_{\rm I}(r)$ and $U_{\rm e}(r)$, acting on ions and electrons, respectively. 
The DF theory can afford to give exact expressions for the 
density distributions $n_i(r|U_{\rm I}U_{\rm e})$ in terms of the inhomogeneous density $n_{i}^{0}({\bf r}|U_{i}^{\rm eff})$ of the noninteracting mixture under the effective external potentials  $U_{i}^{\rm eff}$, which should produce the same density distributions $n_i(r|U_{\rm I}U_{\rm e})$  in such a way \cite{chiharaDF}:
\begin{equation}
n_{i}^{0}({\bf r}|U_{i}^{\rm eff}) \equiv n_{i}({\bf r}|U_{\rm I}U_{\rm e})\,.
\end{equation}
From the above definition, we obtain the explicit expression for  $U_i^{\rm eff}(r)$:
\begin{equation}
U_i^{\rm eff}(r)=U_i(r)\!+\!\delta {\cal F}_{\rm int}/\delta n_i(r)\!-\!\mu_i^{\rm int}
\end{equation}
 in terms of the interaction part of the 
intrinsic free energy ${\cal F}_{\rm int}$ and $\mu_i^{\rm int}\equiv 
\mu_i\!-\!\mu_i^0$. Here, $\mu_{i}$ and $\mu_{i}^0$ are chemical 
potentials of interacting and noninteracting systems of $i$-kind 
particles, respectively. In this way, the DF theory can reduce exactly
 a {\it many}-body problem to determine the density distribution 
$n_{i}({\bf r}|U_{\rm I}U_{\rm e})\,$ of the interacting mixture in the presence 
of the external potential $\{U_{\rm I},U_{\rm e}\}$ to a {\it one}-body problem to
 calculate the density distribution $\,n_{i}^{0}({\bf r}|U_{i}^{\rm 
eff})$ in the noninteracting particles under the effective external 
potential $U_i^{\rm eff}(r)$.

In the second place, with the use of the above result we can express the density distribution of $i$-species particle $n_i(r|\alpha)$, when an $\alpha$-species particle is fixed at the origin ($i$,$\alpha$=I or e):
\begin{equation}
n_i(r|\alpha)=n_{i}^{0}({\bf r}|U_{i\alpha}^{\rm eff})
\end{equation}
with 
\begin{eqnarray}
 U_{i\alpha}^{\rm eff}(r)\!\!&=&\!\!v_{i{\rm \alpha}}(r)+{\delta {\cal F}_{\rm int}\over\delta 
n_i(r|{\rm \alpha})}-\mu_i^{\rm int} \label{e:gvef1}\\   
\!\!&=&\!\!v_{i{\rm \alpha}}(r)-\frac1{\beta}\sum_{\ell}\int C_{i\ell}(|{\bf r}\!-\!{\bf 
r'}|)\,[n_{\ell}}(r'|{\rm \alpha)\!-\!n_0^{\ell}]d{\bf r'}-B_{i{\rm \alpha}}(r)/\beta 
\label{e:gvef2} 
\end{eqnarray}
in terms of the bridge functions $B_{i{\rm \alpha}}(r)$ and the direct 
correlation functions $C_{ij}(r)$. 
In the above expression, the DCFs $C_{ij}(r)$ in the 
ion-electron mixture are defined within the framework of the DF theory \cite{QHNC} 
 by
\begin{equation}
C_{ij}(|{\bf r}-{\bf r'}|) \equiv -\beta { \delta^2 {\cal F}_{\rm 
int}[n_{\rm I},n_{\rm e}]  \over  \delta n_{i}({\bf r})\delta n_{j}({\bf r'}) 
}\biggl |_0  = \beta{\delta [\mu^0_{i}\!\!-\!\!U_{i}^{\rm eff}({\bf 
r})]  \over \delta n_{j}^0({\bf r'}) }\biggl |_0 -\beta{\delta 
[\mu_{i}\!\!-\!\!U_{i}({\bf r})] \over \delta n_{j}({\bf r'}) }\biggl 
|_0 \label{e:dcf} \,,
\end{equation}
where the suffix 0 denotes the functional derivative at the uniform 
densities \cite{c84}.
Actually the explicit expression for the DCFs is given by the 
Fourier transform in the matrix form \cite{JC78}
\begin{equation}
\sqrt{{\cal N}}C(Q)\sqrt{{\cal N}} = (\widetilde{\chi}_Q^0)^{-1} - 
(\widetilde{\chi}_Q)^{-1} \label{e:dcfq} 
\end{equation}
in terms of the density response functions, $\widetilde{\chi}_Q\equiv 
\parallel \chi_{_{ij}}(Q)\parallel$ and $\widetilde{\chi}_Q^0\equiv 
\parallel 
\chi_Q^{0i}\delta_{ij}\parallel$, of the interacting and 
noninteracting 
systems, respectively, with ${\cal N}\equiv \parallel 
n_0^{i}\delta_{ij}\parallel$. 

Here, we assume that ${\cal F}_{\bf Q}[n_i(r|\alpha)\!-\!n_0^i]$ is expressed in terms of 
$\chi_{_{ij}}(Q)$ and the exchange-part of DCF $C_{ii}^{\rm x}(r)$, which works only between the same kind of particles representing the quantum exchange effect:
\begin{equation}\label{e:bootst}
(n_0^in_0^\alpha)^{1/2}{\cal F}_{\bf Q}\left[\frac {n_i(r|\alpha)}{n_0^i}-1 \right]=\frac{\chi_{i\alpha}(Q)}{\chi_Q^{0i} }-\delta_{ij}-\chi_{_{ii}}(Q)n_0^iC_{ii}^{\rm x}(Q)\delta_{i\alpha} \,.
\end{equation}
When the fixed particle is an ion ($\alpha=\rm I$), the above equation provides an exact relation, because of  $\chi_{i\rm I}(Q)=S_{i\rm I}(Q)$, $\chi_Q^{0\rm I}=1$ and $C_{\rm II}^{\rm x}(Q)=0$: Eq.~(\ref{e:bootst}) reduces to the classical relation, $n_i(r|{\rm I})/n_0^i =g_{i\rm I}(r)$, represented by the RDF. Therefore, the following equation only remains as an assumption:
\begin{equation}\label{e:bootstE}
n_0^{\rm e}{\cal F}_{\bf Q}\left[\frac {n_{\rm e}(r|\rm e)}{n_0^{\rm e}}-1 \right]=\frac{\chi_{\rm ee}(Q)}{\chi_Q^{0\rm e} }-1-\chi_{_{\rm ee}}(Q)n_0^{\rm e}C_{\rm ee}^{\rm x}(Q),
\end{equation}
since the fixed electron does not have the exchange-correlation with surrounding electrons.
In the jellium model for an electron gas, we can prove that the exchange part of DCF, $C_{\rm ee}^{\rm x}(Q)$, is related to the exchange part of the local field correction $G_{\rm x}(Q)$, which is 
defined by local field corrections between up- and down-spins, $G_{\uparrow\uparrow}(Q)$ and $G_{\uparrow\downarrow}(Q)$, as follows:
$C_{\rm ee}^{\rm x}(Q)/\beta v_{\rm ee}(Q)= G_{\rm x}(Q)\equiv [G_{\uparrow\uparrow}(Q)-G_{\uparrow\downarrow}(Q)]/2$ \cite{IP,IP2}. 

We can obtain the quantum Ornstein-Zernike relations for this mixture:
\begin{eqnarray}
g_{_{\rm II}}\/(r)-1&=&C_{\rm II}(r)+\Gamma_{\rm II}(r)  \label{e:ozii} 
\\g_{\rm eI}(r)-1
&=&\widehat BC_{\rm eI}(r)+\widehat B\Gamma_{\rm eI}(r) \label{e:ozei}\\
n_{\rm e}(r|{\rm \hat{e}})/n_{\rm e}-1
&=&\widehat B[C_{\rm ee}(r)-C_{\rm ee}^{\rm x}(r)]+
\widehat B \Gamma_{e\hat{e}}(r) \label{e:ozee}
\end{eqnarray}
with
\begin{eqnarray}
\Gamma_{i{\rm I}}(r) &\equiv& \sum_{\ell}\int C_{i\ell}(|{\bf r}-{\bf r'}|) 
\,n_0^{\ell}[g_{\ell {\rm I}}(r')-
1]d{\bf r'} \label{e:gm} \\
\Gamma_{e\hat{e}}(r) &\equiv& \sum_{\ell}\int C_{e\ell}(|{\bf r}-{\bf r}'|)
[n_{\rm \ell}({\bf r}'|{\rm \hat{e}})-n_{\rm e}] d{\bf r}'
\end{eqnarray}
by using a matrix identity
\begin{eqnarray}
&&(\widetilde{\chi}_Q)(\widetilde{\chi}_Q^0)^{-1}-1-\widetilde{\chi}_Q\sqrt{{\cal N}}C^{\rm x}(Q)\sqrt{{\cal N}} \nonumber\\
&& = \widetilde{\chi}_Q^0[(\widetilde{\chi}_Q^0)^{-1}-(\widetilde{\chi}_Q)^{-1}-\sqrt{{\cal N}}C^{\rm x}(Q)\sqrt{{\cal N}}] \nonumber\\ 
&&+ \widetilde{\chi}_Q^0[\widetilde{\chi}_Q^0)^{-1}-(\widetilde{\chi}_Q)^{-1}][(\widetilde{\chi}_Q)(\widetilde{\chi}_Q^0)^{-1}-1-\widetilde{\chi}_Q\sqrt{{\cal N}}C^{\rm x}(Q)]\,.
\end{eqnarray}
Here, $\widehat B$ denotes an operator defined by
\begin{equation}
{\cal F}_{\!\bf Q}[\widehat B f(r)] \equiv \chi_Q^0\!\int e^{i{\bf Q}\cdot{\bf r} }f(r)d{\bf r} \;.
\end{equation} 

In the above, all equations except (\ref{e:bootstE}) are exact, although formal ones, within the framework of the DF theory. 
In an attempt to get a definite expression for the electron-ion interaction 
$v_{\rm eI}(r)$,  we view a plasma more fundamentally as a mixture 
of nuclei and electrons, and obtain $v_{\rm eI}(r)$ in the form \cite{c85}
\begin{equation}
v_{\rm eI}(r)= \tilde v_{\rm eI}(r)  \equiv  -\frac{Z_{\rm A}}{r} +\int v_{\rm ee}(|{\bf 
r}-{\bf 
r'}|)\,\rho_{\rm b}(r')d{\bf r'} +\mu_{\rm XC}(\rho_{\rm b}(r)+n_0^{\rm e}) -\mu_{\rm 
XC}(n_0^{\rm e}) \label{e:vei2} \,, 
\end{equation} 
where $\rho_{\rm b}(r)$ is the bound-electron distribution 
to satisfy $\int \rho_{\rm b}(r)d{\bf r}\!=\!Z_{\rm B}$.
Here, it should be kept in mind that the potential $\tilde v_{\rm eI}(r)$ plays two roles in the same expression: one is the electron-ion interaction $v_{\rm eI}(r)$ under the frozen-core approximation, and the other is the electron-nucleus interaction $v_{\rm eN}(r)$. When $\rho_{\rm b}(r)$ becomes consistent with both the electron-ion model and the AA model, $v_{\rm eI}(r)$ is equal to $v_{\rm eN}(r)$ as mentioned in (II) of \S\ref{s:Sboundfree}.

It is important to note that the above equations for $g_{i\rm I}(r)$ are rewritten as a coupled set of two integral equations: \\
\,\,\,\ (I): One of them is the modified HNC equation to determine the RDF $g_{\rm II}$ for one component fluid interacting via an effective potential $v_{\rm eff}(r)$ \cite{CK1994}:
\begin{equation}\label{eq:QHNCii}
C(r)=\exp[-\beta v_{\rm eff}(r)
  +{\it\gamma}(r)+B_{\rm II}(r)]-1-{\it\gamma}(r)\;,
\end{equation}
with ${\it\gamma}(r)\!\equiv\! \int C(|{\bf r}\!-\!{\bf r}'|)n_0^{\rm I}[g_{\rm II}
(r')-1]d{\bf r}'$ 
 and an interaction $v_{\rm eff}(r)$ defined by 
\begin{equation}\label{eq:veffii}
\beta v_{\rm eff}(Q)\equiv\beta v_{\rm II}(Q)
 -{|C_{\rm eI}(Q)|^2 n_0^{\rm e} \chi_Q^0\over
 1-n_0^{\rm e} C_{\rm ee}(Q) \chi_Q^0 }\;,
\end{equation}
which is described in terms of the pseudopotential $C_{\rm eI}(Q)$.
The above equation (\ref{eq:QHNCii}) is the rewritten form of (\ref{e:ozii})
by introducing the one-component DCF $C(r)$ \cite{CK1994}
\begin{equation}
C(Q)\!\equiv \!\frac{1}{n_0^{\rm I}}\left[ 1\!-\!\frac{1}{S_{\rm II}(Q)} \right]
\!=\!C_{\rm II}(Q)
 \!+\! {|C_{\rm eI}(Q)|^2n_0^e\chi_Q^0 \over 1-n_0^e C_{\rm ee}(Q)\chi_Q^0
 }\!=\! C_{\rm II}(Q)\!+\!C_{\rm eI}(Q)\rho(Q)\,, 
\label{dc}
\end{equation}
which leads to the definition of $\gamma(Q)\!\equiv\!C(Q)[S_{\rm II}-1]\!=\!\Gamma_{\rm II}(Q)\!-\!C_{\rm eI}(Q)\rho(Q)$.

(II): The other is an integral equation (\ref{e:ozei}) to determine the pseudopotential $C_{\rm eI}(Q)$, 
which provides the effective interaction $v_{\rm eff}(r)$ through (\ref{eq:veffii}) to be used in (I):
\begin{equation}\label{eq:QHNCei}
\hat B C_{\rm eI}(r)
 =n_{\rm e}^{\rm f}\left(r|v_{\rm eI}
     -{\it\Gamma}_{\rm eI}/\beta-B_{\rm eI}/\beta\right)/n_0^{\rm e}
     -1-\hat B {\it\Gamma}_{\rm eI}(r)\;. 
\end{equation}
Equation (\ref{eq:QHNCei}) is solved in the AA model, as is discussed by (\ref{e:neutNuc} ).

Here, as mentioned in \S\ref{s:Sboundfree}, the bound-electron distribution $\rho_{\rm b}(r)$ is determined from
\begin{equation}
\rho_{\rm b}(r)=n_{\rm e}(r)-n_{\rm e}^{\rm f}(r)=n_{\rm e}(r)-n_0^{\rm e}g_{\rm eI}(r)=n_{\rm e}(r)-n_0^{\rm e}\!-\!{\cal F}_{\bf r}[\,\rho(Q)S_{\rm II}(Q)\,] \,,
\end{equation}
by using a solution $n_{\rm e}(r)$ of the wave equation under the external potential $U_{\rm eI}^{\rm eff}(r)$ (\ref{e:gvef2}) with $\tilde v_{\rm eI}(r)$ (\ref{e:vei2}). Also, the electron density $n_0^{\rm e}=Z_{\rm I}n_0^{\rm I}$ can be determined by the  condition: 
\begin{equation}\label{e:DfZi2}
\frac{n_0^{\rm e}}{n_0^{\rm I}}
=Z_{\rm A}-\int \Bigl\{ n_{\rm e}(r)\!-\!n_0^{\rm e}\!-\!{\cal F}_{\bf r}[\, \rho(Q)S_{\rm II}(Q)\,]\,\Bigr\} d{\bf r}\!=\!Z_{\rm I} \,,
\end{equation}
or
\begin{equation}\label{e:cndExact2}
\int v_{\rm es}(r)g_{\rm II}(r)d{\bf r}=\mu S_{\rm II}(0)/n_0^{\rm I}=\mu \kappa_{\rm T}/\beta\,.
\end{equation}
The above equation (\ref{e:cndExact2}) is shown to be valid also for the QHNC method if we make further the point-ion or sphere-ion approximation (see Appendix~A).

Here, note that the function $n_{\rm e}^{\rm ext}(r)$, defined to be
$\rho_{\rm pa}(r)\!\equiv\!\rho_{\rm b}(r)\!+\!\rho(r)\!=\!n_{\rm e}(r)\!-\!n_{\rm e}^{\rm ext}(r)$ 
 in the work of Saumon's group, is written as
\begin{equation}
n_{\rm e}^{\rm ext}(r)= n_0^{\rm I}\int \rho (|{\bf r}\!-\!{\bf r}'|)g_{\rm II}(r')d{\bf r}'
\!=\!n_0^{\rm e}g_{\rm eI}(r)\!-\!\rho(r) \,.
\end{equation}
In order to obtain a set of real solutions for above equations, we must introduce some approximations.

%
\section{Variational Average-Atom in Quantum Plasmas (VAAQP) with ion-ion correlation}
%
In the electron-ion model the DCF $C_{\rm eI}(r)$ is related to the relation (\ref{e:ozei}):
\begin{eqnarray}
\hat B C_{\rm eI}(r)
 &=&g_{\rm eI}(r)
     -1-\hat B {\it\Gamma}_{\rm eI}(r) \label{eq:QHNCei2} \,,\\
{\it\Gamma}_{\rm eI}(r) &\equiv&  \sum_l
\int C_{{\rm e}l}(|{\bf r}-{\bf r}'|)n_0^l[g_{l\rm I}(r)-1]d{\bf r}' \,.\label{eq:QHNCgam2}
\end{eqnarray}
At this point, we introduce approximations to $g_{\rm II}(r)$ 
by the step function $\theta (r-R)$ and to 
$C_{\rm eI}(r)\! \simeq\! \beta Z_{\rm I}e^2/r$ in (\ref{eq:QHNCgam2}). These two 
approximations reduce (\ref{eq:QHNCei2}) to an integral equation for the DCF $C_{\rm eI}(r)$, 
since the effective electron-ion interaction $v_{\rm eI}^{\rm eff}(r)$ to determine $g_{\rm eI}(r)$ is written only in terms of $g_{\rm eI}(r)$ without unknown $g_{\rm II}(r)$: 
\begin{eqnarray}
v_{\rm eI}^{\rm eff}(r)&=&\tilde{v}_{\rm eI}(r)+f(r) -\bar{\Gamma}_{\rm 
eI}(r)/\beta \label{JVveff} \,, \\
\tilde{v}_{\rm eI}(r)&=& -\frac{Z_{\rm A}}{r}+\!\int v_{\rm ee}(|{\bf r}\!-\!{\bf r'}|)\rho_{\rm b}(r')d{\bf r}'+\mu_{\rm xc}(\rho_{\rm b}(r)+n_0^{\rm e})\!-\!\mu_{\rm xc}(n_0^{\rm e}) \label{e:Vei} \,,\\
f(r)&\equiv& -\!n_0^{\rm e}\!\!\int \!v_{\rm ee}(|{\bf r}\!-\!{\bf r'}|)[\theta(r'\!\!-\!\!R)\!-\!1]d{\bf r'}     =\left\{ \begin{array}{ll}  Z_{\rm I}[3-(r/R)^2]/2R & 
\mbox{       for $r < R$} \nonumber\\ Z_{\rm I}/r & \mbox{       
for $r \geq  R$} \end{array} \right. \,,
\end{eqnarray}
with $\rho_{\rm b}(r)\equiv n_{\rm e}^{\rm b}(r)+\Delta \rho_{\rm b}(r)$ ; that is, Eq.~(\ref{eq:QHNCei2}) is altered in a simple form (see Reference \cite{QHNa} for details):
\begin{eqnarray}\label{e:JVeq}
\hat B {\bar C}_{\rm eI}(r) = n_{\rm e}^{\rm f}(r|\tilde{v}_{\rm eI}+f-\bar{\Gamma}_{\rm 
eI}/\beta)/n_0^{\rm e}-1-\hat B \bar{\Gamma}_{\rm eI}(r)\,, 
\end{eqnarray}
with the definition of new functions
\begin{equation}\label{e:barGammR}
\bar{\Gamma }_{\rm eI}(r)\equiv \int C_{\rm ee}^{\rm 
jell}(|{\bf r}-{\bf r}'|)n_0^{\rm e}[g_{\rm eI}(r)-1]{\rm d}{\bf r}'\,, 
\end{equation}
\begin{equation}
\bar{C}_{\rm eI}(r)\equiv C_{\rm eI}(r)+n_0^{\rm e}\int \beta 
\frac{e^2}{|{\bf r}-{\bf r}'|}[\theta(r'-R)-1]{\rm d}{\bf r}'=C_{\rm eI}(r)-\beta f(r)\,.
\end{equation}
Here, $\bar{C}_{\rm eI}(Q)$ is related to $\bar{\Gamma }_{\rm eI}(Q)$ by the relation:
\begin{equation}
\bar{\Gamma}_{\rm eI}(Q)={n_0^{\rm e} \chi_{\rm ee}^{\rm jell}C_{\rm ee}^{\rm 
jell}(Q)}\bar{C}_{\rm eI}(Q) \label{e:barGamm}\,.
\end{equation}
and the suffix 'jell' denotes a corresponding quantity in the jellium model. 

Thus, equation (\ref{e:JVeq}) with (\ref{e:barGamm}) becomes an integral equation for 
$\bar{C}_{\rm eI}(r)$ in conjunction with $n_{\rm e}^{\rm f}(r)\equiv
n_{\rm e}(r)\!-\!n_{\rm e}^{\rm b}(r)\!-\!\Delta\rho_{\rm b}(r)\!=\!n_0^{\rm e}g_{\rm eI}(r)
$, 
since $\bar{\Gamma}_{\rm eI}(r)$ is expressible in terms of 
$\bar{C}_{\rm eI}(r)$ by means of (\ref{e:barGamm}). Thus, $\bar{C}_{\rm eI}(Q)$ obtained from a set of integral equations yields a pseudo-potential ${C}_{\rm eI}(Q)$ in the form:
\begin{eqnarray}\label{e:JVcei}
C_{\rm eI}(Q)&=&{\bar C}_{\rm eI}(Q)+\beta f(Q) \,,\\
f(Q)&\equiv& n_0^{\rm e}{\cal F}_Q[f(r)]=3Z_{\rm I}\frac{\sin(RQ)-(RQ)\cos(RQ)}{(RQ)^3}\,.
\end{eqnarray}
Eq.~(\ref{e:JVcei}) indicates that the pseudo-potential $C_{\rm eI}(r)/(-\beta)$ in large distances behaves as $-f(r)\!=\!-Z_I/r\!=\!-n_0^{\rm e}/(n_0^{\rm I}r)$ for a given $n_0^{\rm e}$. 
In this way, when the DCF $C_{\rm eI}(Q)/(\!-\!\beta$) is known, the effective ion-ion potential $v_{\rm eff}(r)$ can be calculated by the formula
\begin{equation}\label{eq:veffii2}
\beta v_{\rm eff}(Q)\equiv\beta v_{\rm II}(Q)
 -|C_{\rm eI}(Q)|^2 n_0^{\rm e}\chi_Q^{jell} \,.
\end{equation}
Once the interaction $v_{\rm eff}(r)$ is given, the ion-ion RDF $g_{\rm II}(r)$ is obtained by solving the HNC integral equation for a fluid:
\begin{equation}\label{eq:QHNCii20}
C(r)=\exp[-\beta v_{\rm eff}(r)
  +{\it\gamma}(r)]-1-{\it\gamma}(r) \,.
\end{equation}
Therefore, even in this model where $g_{\rm II}(r)$ is approximated by the step function, we can use the RDF $g_{\rm II}(r)$ to determine the electron density $n_0^{\rm e}$ from the condition:
\begin{equation}\label{e:selfn0}
\int \left.\frac{\delta E_{\rm es}}{\delta n_{\rm e}({\bf r})}\right|_{U(\bf r)}\!\!g_{\rm II}(r)d{\bf r}=\mu S_{\rm II}(0)/n_0^{\rm I}=\mu \kappa_{\rm T}/\beta \,,
\end{equation}
where 
\begin{equation}
\left.\frac{\delta E_{\rm es}}{\delta n_{\rm e}({\bf r})}\right|_{U({\bf r})}
= U(r)+\int v_{\rm ee}(|{\bf r}\!-\!{\bf r'}|)[n_{\rm e}(r')\!-\!n_0^{\rm e}]d{\bf r}' \,. \label{e:dEesDdn3} 
\end{equation}

In the above, we pay attention to the exchange-correlation effect $\mu_{\rm xc}({\bf r}|n_{\rm e})$ in Eq.~(\ref{e:vei2}), where it is described in terms of $C_{\rm ee}^{\rm xc}\!\equiv\! C_{\rm ee}\!+\!\beta v_{\rm ee}$ as
\begin{equation}\label{e:CeeX}
\mu_{\rm xc}({\bf r}|n_{\rm e})\approx \mu_{\rm xc}(\rho_{\rm b}(r)\!+\!n_0^{\rm e})-\mu_{\rm xc}(n_0^{\rm e}) 
+\int \frac{C_{\rm ee}^{\rm xc}(|{\bf r}-{\bf r}'|)}{-\beta}[n_{\rm e}^{\rm f}(r')-n_0^{\rm e}]d{\bf r}' \,.
\end{equation}
On the other hand, if we use the LDA, $\mu_{\rm xc}({\bf r}|n_{\rm e})\!\approx\! \mu_{\rm xc}(n_{\rm e}({\bf r}))\!-\!\mu_{\rm xc}(n_0^{\rm e})$, instead of (\ref{e:CeeX}), the effective interaction (\ref{JVveff}) for electrons becomes 
\begin{equation}
{v}_{\rm eI}^{\rm eff}(r) =-\frac{Z_{\rm A}}{r}+f(r)+\int v_{\rm ee}(|{\bf r}\!-\!{\bf r'}|)
[n_{\rm e}(r')\!-\!n_0^{\rm e}]d{\bf r}'+\mu_{\rm xc}(n_{\rm e}({\bf r}))\!-\!\mu_{\rm xc}(n_0^{\rm e}) \label{JVveffLDA} \,,
\end{equation}
which is commonly used in the NPA model. Since Eq.~(\ref{JVveffLDA}) is represented only by $n_{\rm e}({\bf r})$, any separation of $\rho_{\rm b}(r)$ from $n_{\rm e}({\bf r})$ is unnecessary to evaluate this effective interaction.
With the use of this approximation, when $n_{\rm e}(r)\!=\!\rho_{\rm b}^{[\alpha\!-\!1]}(r)\!+\!n^{\rm f{[\alpha\!-\!1]}}_{\rm e}(r)$ and $n_0^{\rm e}$ are given, we can 
determine the next free-electron distribution $n^{\rm f{[\alpha]}}_{\rm e}(r)$ in the iteration, and  obtain the new DCF for calculating new $S_{\rm II}(Q)$ as follows:

In the frozen-core approximation where $\rho_{\rm b}^{[\alpha\!-\!1]}(r)$ is treated as fixed, the wave equation can be solved to determine $n^{\rm f{[\alpha]}}_{\rm e}(r)\!=\!n_0^{\rm e}g_{\rm eI}(r)$ under the external potential $v_{\rm eI}^{\rm eff}(r)$. In this calculation, as a free-electron contribution we must take account of the bound states in $n_{\rm e}^{\rm b}(r)$, which appeared as a shallow bound state according to the temperature increase.
 The total electron distribution  $n_{\rm e}(r)$ with the fixed electron density $n_0^{e}$ is invariant with respect to the free- and bound-electron division if we use the approximation (\ref{JVveffLDA}). Therefore, new bound-electron distribution is obtained by $\rho_{\rm b}^{[\alpha]}(r)\!=\!n_{\rm e}(r)\!-\!n^{\rm f{[\alpha]}}_{\rm e}(r)$. In this way, we can divide the electron distribution $n_{\rm e}(r)$ in the bound and free parts self-consistently. Thus, we can obtain new DCF to determine new $S_{\rm II}(Q)$ from Eq.~(\ref{e:JVeq}).
For simple metals, this set of equations (referred to as NPA model) is easily solved because of $\Delta\rho_{\rm b}(r)\!=\!0$, and the structure factor $S_{\rm II}(Q)$ calculated by this  model shows an excellent agreement with the experiments \cite{gonzalk,gonzearth}, while the RDF $n_0^{\rm e}g_{\rm eI}(r)$, when determined by $n_0^{\rm e}\!+\!{\cal F}_{\bf r}[\, \rho(Q)S_{\rm II}(Q)]$, agrees with the results from the QHNC method \cite{cdwp,npa,gonzalk,gonzearth}.

An essential point of the VAAQP method is based on the jellium model as an approximate model to the electron-ion model to treat the electron system. Therefore, the VAAQP method can provide only the electron pressure $P_{\rm e}$. We have proven that the pressure $P$ of a plasma is given by the sum of electron pressure and nuclear pressure by regarding a plasma as an electron-nucleus mixture in the form (\, (4.2) in Reference \cite{chiharaNe}, (7.2) in Reference \cite{chiharaDF} and (91) in Reference \cite{chi2001} \, ):
\begin{eqnarray}\label{e:flEOS1}
P&=&P_{e}+n_0^{\rm I}k_{\rm B}T
-\frac16 (n_0^{\rm I})^2 \!
\int g_{\rm II}(r)\,{\bf r}\cdot\nabla v_{\rm eff}(r)d{\bf r} \,.
\end{eqnarray}
Note that the VAAQP method gives all quantities in the above equation.
Equation (\ref{e:flEOS1}) is derived on an assumption that the force ${\bf F}_\alpha$ on $\alpha$-nucleus in the plasma, produced by electron clouds and surrounding nuclei, is the same to the force on the $\alpha$-pseudo-atom by surrounding pseudo-atoms (average atoms) in the electron-ion model: 
 ${\bf F}_\alpha\!=\!-\sum_\beta \nabla_\alpha v_{\rm eff}(|{\bf R}_\alpha\!-\!{\bf R}_\beta |)$
. 
Relating to (\ref{e:flEOS1}), we have replied  in our notes \cite{chiharaLPP} a question that the law of partial pressures is applicable to an interacting system.

%
\section{QHNC equation for an electron gas in the uniform jellium}
%

In order to determine the electron-electron effective interaction $v_{\rm ee}^{\rm eff}$ in an electron gas, Kukkonen and Overhauser \cite{KO} have made an assumption that the accumulated electrons $n_{\rm e}(r|{\rm e})$ around the electron in an electron gas are given by the electron density distribution $n_{\rm e}(r|{\rm e})$ around the fixed electron in an electron gas, and the second assumption that the effective potential between an infinitesimal small test charge $\delta e$ and the electron in an electron gas should have the following relation
\begin{equation}\label{e:AnsatzKO}
V_{\rm \delta e, \widehat{e}}^{\rm eff}
=V_{\rm e, \widehat{\delta e}}^{\rm eff}\,.
\end{equation}
Here, $V_{\rm \delta e, \widehat{e}}^{\rm eff}$ is an effective potential between the test charge and the fixed electron, and $V_{\rm e, \widehat{\delta e}}^{\rm eff}$ is a potential for the electron in an electron gas caused by the fixed test charge.  These assumptions lead to the following equation\cite{JC78EG}:
\begin{equation}\label{e:KOappr2}
{\cal F}_Q[n_{\rm e}(r|{\rm {e}})-n_{\rm e}]
=\chi_{\rm ee}(Q)/\chi_Q^0-1\,.
\end{equation}
Equation~(\ref{e:KOappr2}) involves an approximation that the "fixed electron" in an electron gas has the exchange-effect on surrounding electrons.  

To see the meaning of the "fixed electron", let us put two test charges in the electron gas: 
the one has an infinitesimal charge $\delta e$ and the mass $m_{\delta e}$
and the other T with a finite charge $e_{\rm T}$ and the mass $m_{\rm T}$. 
The electrostatic potential for the large test charge T arising from the fixed test charge 
$\delta e$ should be equal to the electrostatic potential 
for the small test charge $\delta e$ produced by the fixed large test charge T, since the electrostatic potential between two particles does not depend on their masses:
\begin{equation}\label{e:electrost}
V_{\rm electrostatic}\equiv V_{\rm \delta e, \widehat{e}}^{\rm eff}
=V_{\rm e, \widehat{\delta e}}^{\rm eff}\,.
\end{equation}
From this relation with $e_{\rm T}\!=\!e$, we have a bootstrap relation as follows, 
\begin{eqnarray}
{\cal F}_Q[n_{\rm e}(r|{\rm \hat{e}})-n_{\rm e}]
&=&\chi_{\rm ee}(Q)/\chi_Q^0-1-n_{\rm e}C_{\rm ee}^{\rm x}(Q)\chi_{\rm ee}(Q)\label{e:BootSt}\\
&=&\chi_{\rm ee}(Q)/\chi_Q^0-1-n_{\rm e}\beta v_{\rm ee}(Q)G_{\rm x}(Q)\chi_{\rm ee}(Q) \,,
\end{eqnarray}
and the Ornstein-Zernike equation for $n_{\rm e}(r|{\rm \hat{e}})/n_{\rm e}$ is written as
\begin{equation}\label{e:OZx2}
n_{\rm e}(r|{\rm \hat{e}})/n_{\rm e}-1=
\widehat B[C_{\rm ee}(r)-C_{\rm ee}^{\rm x}(r)]+\widehat B\int C_{\rm ee}(|{\bf r}-{\bf r}'|)
[n_{\rm e}({\bf r}'|{\rm \hat{e}})-n_{\rm e}] d{\bf r}' \,.
\end{equation}
However, we have no integral equation to determine $C_{\rm ee}^{\rm x}(r)$. Since the exchange DCF $C_{\rm ee}^{\rm x}(r)$ has a very short-range effect, it may be allowed to approximate the LFC $G_{\rm x}(Q)$ by that of the jellium model. In treating a hydrogen liquid, the exchange part $G_{\rm x}(Q)$  is important, and the correlation part $G_{\rm c}(Q)$ should be determined in a coupled manner with the ion-structure \cite{JC84,chiharaCMPhyd}.

%
\section{Discussion}
%
The definition of the ionization state $Z_{\rm I}$ itself is subject to controversy
because this quantity is not the eigenvalue of any operator \cite{npa}.
However, the DF theory ensures that the electron density distribution $n_{\rm e}(r)$  is an observable physical quantity, although the bound- and the continuum-electron distributions, $n_{\rm e}^{\rm b}(r)$ and $n_{\rm e}^{\rm c}(r)$, are not ensured separately as physical quantities.
On the other hand, in an ion-electron mixture the ion-electron and ion-ion RDFs are also physical quantities ensured by the DF theory, when the interactions, $\tilde v_{\rm eI}(r)$ and $v_{\rm II}(r)$, are known. Based on this view, we defined the 'ion' in the AA model in such a way that the 'ion'-electron RDF $n_{\rm e}^{\rm f}(r)$ around this 'ion' becomes identical to the ion-electron RDF in the ion-electron mixture in the form (\ref{e:nef}):
\begin{equation}\label{e:nef2}
n_{\rm e}(r)=n_{\rm e}^{\rm b}(r)\!+\!n_{\rm e}^{\rm c}(r)=\rho_{\rm b}(r)\!+\!n_{\rm e}^{\rm f}(r)=\rho_{\rm b}(r)\!+\!n_0^{\rm e}g_{\rm eI}(r) \,.
\end{equation}
Therefore, the ionization state $Z_{\rm I}$ is a physical quantity defined in terms of physical quantities, $g_{\rm eI}(r)$ and $n_{\rm e}(r)$. Here, the electron-ion RDF $g_{\rm eI}(r)$ and the density $n_0^{\rm e}$ can be determined by the QHNC equations, (\ref{eq:QHNCii}) and (\ref{eq:QHNCei}), self-consistently  
for an electron-ion mixture interacting via $v_{\rm II}(r)$ and $\tilde v_{\rm eI}(r)$ with the frozen-core $\rho_{\rm b}(r)$, in accordance with the relation (\ref{e:nef2}): here, the interaction $v_{\rm II}(r)$ assumed as known is determined as consistent with $\rho_{\rm b}(r)$.

To see the meaning of $Z_{\rm I}$ as a physical quantity, let us consider an impurity problem where an ion $Z_{\rm I}$ is immersed in the jellium at density $n_0^e$ and temperature $T$, for simplicity.
If the ion-electron interaction $\tilde v_{\rm eI}(r)$ is given beforehand, the DF theory provides an integral equation to determine the electron-ion RDF around the fixed ion under the QHNC approximation: 
\begin{eqnarray}
n_0^e g_{\rm eI}(r)&=& n^0_{\rm e}(r|\tilde{v}_{\rm eI}\!-\!{\Gamma}_{\rm eI}/\beta)\,,\label{e:geiImp}\\
{\it\Gamma}_{\rm eI}(r) &\equiv&  
\int C_{\rm ee}^{\rm jell}(|{\bf r}-{\bf r}'|)n_0^e[g_{\rm eI}(r)-1]d{\bf r}' \,\label{eq:QHNCgamJel}\,,\\
\hat B C_{\rm eI}(r) &=&g_{\rm eI}(r)
     -1-\hat B {\it\Gamma}_{\rm eI}(r) \label{eq:QHNCeiJel} \,.
\end{eqnarray}
Here, $n^0_{\rm e}(r|v_{\rm eI}^{\rm eff})$ is calculated by solving a wave equation under the external potential $v_{\rm eI}^{\rm eff}(r)\!=\!\tilde{v}_{\rm eI}(r)\!-\!{\Gamma}_{\rm eI}(r)/\beta$. 
This set of equations becomes an integral equation also to determine the electron-nucleus RDF $g_{\rm eN}(r)$ as another impurity problem fixed a nucleus $Z_{\rm A}$ in the jellium, if we adopt Eq.~(\ref{e:vei2}) as $\tilde{v}_{\rm eI}(r)$ with the use of division (\ref{e:nef2}). 
The charge neutrality condition for this problem is written as
\begin{eqnarray}
Z_{\rm A}\!\!\!&=&\!\!\!\!\int [n_{\rm e}(r)\!-\!n_0^{\rm e}]d{\bf r}=\!\!\!\sum_{\epsilon_i<0}\!f(\epsilon_i)\!+\!\frac2{\pi}\sum_{\ell}
(2\ell\!+\!1)\,\!\int_0^{\infty}\!\!\!f(E){d\delta_{\ell}(E)\over dE}dE \nonumber\\
&=&\!\!\!\sum_{\epsilon_i<0}\!f(\epsilon_i)\!+\!\Delta Z_{\rm B}+Z_{\rm I}\,,
\end{eqnarray}
by defining the number of bound electron: 
\begin{equation}
Z_{\rm B}\!=\!\int [n_{\rm e}(r)\!-\!\!n_0^{\rm e}g_{\rm eI}(r)]d{\bf r}\!=\!\int \rho_{\rm b}(r)d{\bf r}\!\!=\!\!\sum_{\epsilon_i<0}\!f(\epsilon_i)\!+\!\Delta Z_{\rm B}\,,
\end{equation}
with $Z_{\rm I}\equiv\int [n_0^{\rm e}g_{\rm eI}(r)\!-\!n_0^{\rm e}]d{\bf r}$ and $n_0^{\rm e}g_{\rm eI}(r)\!=\!n_{\rm e}^{\rm c}(r)-\Delta \rho_{\rm b}(r)$.
Here, the electron-ion RDF $g_{\rm eI}(r)$ is calculated from Eq.~(\ref{e:geiImp}) with the use of the electron-ion interaction $\tilde{v}_{\rm eI}(r)$ where the core electron $\rho_{\rm b}(r)$ is taken to be frozen; when the electron-ion RDF $g_{\rm eI}(r)$ is determined self-consistently to fulfill the division (\ref{e:nef2}), then, the bound-electron distribution $\rho_{\rm b}(r)$ and $Z_{\rm I}$ become physical quantities in contrast with the bound-electron density $n_{\rm e}^{\rm b}(r)$. 
The first example is a proton impurity in the jellium at zero temperature. There is no bound state when the electron sphere radius $r_{\rm s}$ is smaller than 2. For larger values of $r_{\rm s}$ than 2 the 1s-bound state appears, and deepens and forms a H$^{-}$ ion $n_{\rm e}^{\rm b}(r)$ in accord with increase of $r_{\rm s}$ \cite{Zar1977,Almb1976}. However, we cannot take the H$^{-}$ ion as $\rho_{\rm b}(r)$, since the RDF $n_0^eg_{\rm eI}(r)\!=\!n_{\rm e}^{\rm c}(r)$ has a negative part \cite{Manninen1981}; in reality, we have the RDF $n_0^{\rm e}g_{\rm ep}(r)\!=\!n_{\rm e}^{\rm b}(r)\!+\!n_{\rm e}^{\rm c}(r)$  with $\rho_{\rm b}(r)\!=\!0$ for all $r_{\rm s}$, that is $Z_{\rm I}\!=\!1$. 
In the case of Al impurity at $r_{\rm s}\!=\!2.07$, there appears a weak 3s$^2$ bound state, whose density is added to the continuum density $n_{\rm e}^{\rm c}(r)$ to get $g_{\rm eI}(r)$ in the determination of $Z_{\rm I}\!=\!3$, as is the same to the assumption made by Manninen {\it et al} \cite{Manninen1981}.

In a proton-electron mixture which is in a liquid metal state without any bound state [$n_{\rm e}^{\rm b}(r)\!\!=\!\!\rho_{\rm b}(r)\!\!=\!\!0$], the 1s-level appears when the temperature is increased at constant density \cite{Perrot1984}: even in this state, the electron-proton RDF is given by 
$n_0^{\rm e}g_{\rm ep}(r)\!=\!n_{\rm e}^{\rm b}(r)\!+\!n_{\rm e}^{\rm c}(r)$ because of $\rho_{\rm b}(r)\!\!=\!\!0$ in a similar manner as a hydrogen plasma gas \cite{ChiharaHplasma}.
In the case of a rubidium plasma \cite{PR99plas}, 
the shallow bound levels are shown to appear with increase of the temperature in fig.~10, and the electron-ion RDF without involving these shallow bound states does not exhibit the continuous temperature-variation after the appearance of the shallow 4d-level in fig.~7. Such a shallow-level density should be involved in the free-electron distribution as is done for a beryllium plasma \cite{Plag2015}.
The size of ion $\rho_{\rm b}(r)$  becomes as smaller as the temperature increases, 
owing to the subtraction of the shallow-level density from $n_{\rm e}^{\rm b}(r)$.
In the thermal ionization, the charge $Z_{\rm A}\!-\!\int n_{\rm e}^{\rm b}(r)d{\bf r}\!-\!n_{\rm 0}^{\rm e}/n_{\rm 0}^{\rm I}$ becomes negative, if there is no resonant state. 

%
                  \appendix
%

\section{The central ion as an atom in the AA model}
Let us consider 
a chosen central ion as an atom immersed in an electron-ion mixture with a nucleus fixed at the origin of coordinates. The central nucleus accumulates electrons $n_{\rm e}(r|{\rm N})$, and pushes away the surrounding ions $n_{\rm I}(r|{\rm N})$. The free energy of this system is written as
\begin{equation}\label{e:exactFE}
F[n_{\rm e}(r|{\rm N}),n_{\rm I}(r|{\rm N})]={\cal F}[n_{\rm e},n_{\rm I}]
\!+\!\!\int v_{\rm eN}(r)n_{\rm e}(r|{\rm N})d{\bf r}\!+\!\!\int v_{\rm IN}(r)n_{\rm I}(r|{\rm N})d{\bf r}\,,
\end{equation}
when the fixed nucleus causes external potentials, $v_{\rm eN}(r)$ and $v_{\rm IN}(r)$, for electrons and ions, respectively, in terms of the intrinsic free energy ${\cal F}[n_{\rm e},n_{\rm I}]$.
In the point-ion approximation which provides $E_{\rm es}$ by (\ref{e:Ees}), this free energy is represented as
\begin{equation}\label{e:piFE}
F[n_{\rm e}(r|{\rm N}),n_{\rm I}(r|{\rm N})]={\cal F}_{\rm s}[n_{\rm e},n_{\rm I}]+{\cal F}_{\rm xc}[n_{\rm e},n_{\rm I}]+E_{\rm es}\,,
\end{equation}
with 
\begin{equation}
{\cal F}_{\rm s}=\frac1{\beta}\int n_{\rm I}({r}|{\rm N})\ln[n_{\rm I}({r}|{\rm N})\lambda^3]d{\bf r} +{\cal F}_{\rm s}^{\rm e}[n_{\rm e}({\bf 
r}|{\rm N})] \label{e:f0} \,,
\end{equation}
where ${\cal F}_{\rm s}^{\rm e}[n_{\rm e}({r}|{\rm N})]$ is the intrinsic free energy of 
noninteracting electrons, and $\lambda$ indicates the thermal 
wavelength of ion.  
When the ion-ion interaction is represented by $v_{\rm II}(r)\!=\!v_{\rm II}^0(r)\!+\!Z_{\rm I}^2/r$  instead of a point ion,
the point-ion free energy (\ref{e:piFE}) is only changed by an additional term $\int \!v_{\rm II}^0(r) n_{\rm I}(r|{\rm N})d{\bf r}$ in the external potential caused by the central nucleus as its free energy: 
\begin{equation}\label{e:piFEv0}
F[n_{\rm e}(r|{\rm N}),n_{\rm I}(r|{\rm N})]={\cal F}_{\rm s}[n_{\rm e},n_{\rm I}]+{\cal F}_{\rm xc}[n_{\rm e},n_{\rm I}]+E_{\rm es}+\int \!v_{\rm II}^0(r) n_{\rm I}(r|{\rm N})d{\bf r}\,.
\end{equation}
From the above equation the effective ion-nucleus interaction 
$v_{\rm IN}^{\rm eff}({r})$ is derived as follows. At first, because of $n_{\rm I}({\bf r}|{\rm N})\!=\!n_{\rm I}({r}|{\rm I})$ and $g_{\rm eN}({r})\!=\!g_{\rm eI}({r})$, we obtain 
\begin{eqnarray}
\frac{\delta E_{\rm es}}{\delta n_{\rm I}({r}|{\rm N})}\!=\!\frac{\delta E_{\rm es}}{\delta [n_0^{\rm I}g(r)]}=\frac{Z_{\rm I}Z_{\rm A}}{r}\!-\!Z_{\rm I}\!\int \!v_{\rm ee}(|{\bf r}\!-\!{\bf r}'|)\{n_{\rm e}(r')\!-\!n_0g(r') \}d{\bf r}' 
\end{eqnarray}
Second, we apply the frozen-core approximation to $n_{\rm e}({\bf r}|{\rm N})\!=\!\rho_{\rm b}(r)\!+\!n_0g_{\rm e}({\bf r}|{\rm N})$, there results
\begin{eqnarray}
&&\left.\frac{\delta{\cal F}_{\rm xc}[\rho_{\rm b}\!+\!n_{\rm e}^{\rm f},n_{\rm I}]}{\delta n_{\rm I}({\bf r}|{\rm N})}\right|_{\rho_{\rm b}:{\rm fixed}}\!\!\!\approx \frac{\delta{\cal F}_{\rm xc}[n_{\rm e}^{\rm f},n_{\rm I}]}{\delta n_{\rm I}({\bf r}|{\rm I})}\nonumber\\
=&&\!\!\!\!\!\!\!\!n_0\!\int \!\frac{ C_{\rm Ie}^{\rm xc}(|{\bf r}\!\!-\!\!{\bf r}'|)}{-\beta}\{ g_{\rm eI}({r'})\!-\!1 \}d{\bf r}'
 +n_0^{\rm I}\!\int \!\frac{ C_{\rm II}^{\rm xc}(|{\bf r}\!\!-\!\!{\bf r}'|)}{-\beta}\{ g({r'})\!-\!1 \}d{\bf r}' \,.
\end{eqnarray}
The frozen-core approximation is a fundamental assumption to build up the electron-ion model 
for an electron-nucleus mixture.
In final, the effective ion-ion interaction $v_{\rm IN}^{\rm eff}({r})$ is obtained as
\begin{eqnarray}
v_{\rm IN}^{\rm eff}({r})\!&=&\! v_{\rm IN}({r})\!+\!{\delta {\cal F }_{\rm int}[n_{\rm e},n_{\rm I}]}/{\delta n_{\rm I}({\bf r}|{\rm N})} \\
=&&\!\!\!\!\!\!\!\!\!\!\!\!\!v_{\rm IN}({r})\!+\!\!n_0\!\int \!\frac{ C_{\rm Ie}(|{\bf r}\!\!-\!\!{\bf r}'|)}{-\beta}\{ g_{\rm eI}({r'})\!-\!1 \}d{\bf r}'
 +n_0^{\rm I}\!\int \!\frac{ C_{\rm II}(|{\bf r}\!\!-\!\!{\bf r}'|)}{-\beta}\{ g({r'})\!-\!1 \}d{\bf r}' \,,\nonumber
\end{eqnarray}
with ${\cal F}_{\rm int}\!\equiv\!{\cal F}\!-\!{\cal F}_{\rm s}$ and
\begin{eqnarray}
v_{\rm IN}({r})&=&v_{\rm II}^0(r)+\frac{Z_{\rm I}Z_{\rm A}}{r}\!+\!\int \!v_{\rm ee}(|{\bf r}\!-\!{\bf r}'|)\rho_{\rm b}(r')d{\bf r}' \nonumber\\
&=& v_{\rm II}^0(r)+\frac{Z_{\rm I}^2}{r}=v_{\rm II}(r)\,,
\end{eqnarray}
since two ions do not approach each other in the distance smaller than a repulsive core diameter $\sigma$ of $v_{\rm II}^0(r)$: $\int_0^\sigma \rho_{\rm b}(r)d{\bf r}\approx Z_{\rm B}$, that is,
\begin{equation}
\frac{Z_{\rm A}}{r}-\!\int\! v_{\rm ee}(|{\bf r}\!-\!{\bf r}'|)\rho_{\rm b}(r')d{\bf r}'
=\frac{Z_{\rm A}\!\!-\!\!\int_0^r\rho_{\rm b}(r')d{\bf r}'}{r}-\int_r^\infty\frac{\rho_{\rm b}(r')}{r'}d{\bf r}'\approx \frac{Z_{\rm I}}{r} \,.
\end{equation}
Thus, the effective ion-nucleus interaction $v_{\rm IN}^{\rm eff}({r})$ and $v_{\rm IN}({r})$ are shown to be identical with $v_{\rm II}^{\rm eff}({r})$ and $v_{\rm II}({r})$ of the electron-ion model, respectively; the condition to define the "ion" is fulfilled under these approximations. In addition, the effective electron-nucleus interaction $v_{\rm eN}^{\rm eff}({r})$ is given by (3.8) in reference \cite{QHNC}.  Due to the relation ${\delta F}/{\delta n_{\rm e}({\bf r}|{\rm N})}\!=\!\mu_{\rm e}[n_0,n_0^{\rm I}]$, we obtain (\ref{e:cndExact}) by following the same procedure as described in \S\ref{JVion}, since it is derived from the same condition (\ref{e:n0C}).

\section{The thermal ionization}
When a new bound state $\epsilon_{\rm b}$ with a density $n_{\rm e}^{\epsilon_{\rm b}}(r)$ appears in accompany with the temperature increase, 
the continuum-electron density $n_{\rm e}^{\rm c}(0)$ and the bound-electron density $n_{\rm e}^{\rm b}(0)$ at the origin 
have a cusp singularity around this temperature, although the total electron distribution $n_{\rm e}(0)$ 
has no singularity \cite{Cusp}. Note that the bound-electron distribution $\rho_{\rm b}(r)\!=\!n_{\rm e}^{\rm b}(r)\!-\!n_{\rm e}^{\epsilon_{\rm b}}(r)$ and the free-electron distribution $n_{\rm e}^{\rm f}(r)\!=\!n_{\rm e}^{\rm c}(r)\!+\!n_{\rm e}^{\epsilon_{\rm b}}(r)$ have no singularity as well as $n_{\rm e}(r)$. This fact is reasonable since this bound level $\epsilon_{\rm b}$ has no physical meaning, as is seen for the case of a hydrogen plasma gas, which keeps a fully ionized state even when changed to a liquid metal hydrogen with a shallow 1s-level \cite{ChiharaHplasma,JC84}.

\providecommand{\noopsort}[1]{}\providecommand{\singleletter}[1]{#1}%

\end{document}